\documentclass[11pt]{article}
\usepackage{epsfig}
\usepackage{enumerate}

\def\be{\begin{equation}}
\def\ee{\end{equation}}
\def\bea{\begin{eqnarray}}
\def\eea{\end{eqnarray}}
\newcommand{\lsim}{\mathrel{\mathop{\kern 0pt \rlap
  {\raise.2ex\hbox{$<$}}}
  \lower.9ex\hbox{\kern-.190em $\sim$}}}
\newcommand{\gsim}{\mathrel{\mathop{\kern 0pt \rlap
  {\raise.2ex\hbox{$>$}}}
  \lower.9ex\hbox{\kern-.190em $\sim$}}}

\newcommand{\AmS}{{\protect\the\textfont2
  A\kern-.1667em\lower.5ex\hbox{M}\kern-.125emS}}
\normalsize
\begin{document}

\baselineskip=0.65cm

\begin{center}
\Large
{\bf A few final comments to arXiv:1210.7548[hep-ph]}
\vspace{0.5cm}

\rm
\end{center}

\large

\begin{center}

R.\,Bernabei$^{1,2}$,~P.\,Belli$^{2}$,~F.\,Cappella$^{3,4}$,~V.\,Caracciolo$^{5}$,~R.\,Cerulli$^{5}$,
\vspace{1mm}

C.J.\,Dai$^{6}$,~A.\,d'Angelo$^{3,4}$,~A.\,Di Marco$^{1,2}$,~H.L.\,He$^{6}$,~A.\,Incicchitti$^{4}$,
\vspace{1mm}

X.H.\,Ma$^{6}$,~F.\,Montecchia$^{2,7}$,~X.D.\,Sheng$^{6}$,~R.G.\,Wang$^{6}$ and Z.P.\,Ye$^{6,8}$
\vspace{1mm}

\normalsize

\vspace{0.4cm}

$^{1}${\it Dip. di Fisica, Universit\`a di Roma ``Tor Vergata'', I-00133 Rome, Italy}
\vspace{1mm}

$^{2}${\it INFN, sez. Roma ``Tor Vergata'', I-00133 Rome, Italy}
\vspace{1mm}

$^{3}${\it Dip. di Fisica, Universit\`a di Roma ``La Sapienza'', I-00185 Rome, Italy}
\vspace{1mm}

$^{4}${\it INFN, sez. Roma, I-00185 Rome, Italy}
\vspace{1mm}

$^{5}${\it Laboratori Nazionali del Gran Sasso, I.N.F.N., Assergi, Italy}
\vspace{1mm}

$^{6}${\it IHEP, Chinese Academy, P.O. Box 918/3, Beijing 100039, China} 
\vspace{1mm}

$^{7}${\it Laboratorio Sperimentale Policentrico di Ingegneria Medica, Universit\`a
degli Studi di Roma ``Tor Vergata''}
\vspace{1mm}

$^{8}${\it University of Jing Gangshan, Jiangxi, China}
\vspace{1mm}

\end{center}

\normalsize

\begin{abstract}
\noindent 
A few final comments on arXiv:1210.7548 are given to confute incorrect arguments claimed 
there.
\end{abstract}

After our comment \cite{replay1} on the arguments of ref. \cite{scem},
some of the same authors persist in some misleading arguments \cite{scem2}.
Here, as our last remarks on the topic, we comment a few points with particular emphasis to 
the arguments related to the extraction of the Dark Matter signal 
through the annual modulation signature and the role of the modeling of the 
background components in the {\it single-hit} counting rate.

\section{Extraction of the Dark Matter signal through the annual modulation signature}

Firstly, let us briefly recall a few arguments in order to focus the point.
Two different approaches are exploited in the field of Dark Matter (DM) direct detection. 
The first one aims at extracting the constant part (S$_0$) of the signal
from the measured counting rate, assuming a particular class of DM candidates. 
In experiments exploiting such an approach
many kinds of uncertain subtractions/selections of the measured events are applied to select 
recoil-like candidates.

The second approach, followed by the DAMA experiments, exploits a model-independent signature 
with very peculiar features: the DM annual modulation signature (see for example \cite{modlibra,modlibra2}). 
In this case the experimental observable is not S$_0$, but the modulation amplitude, S$_m$, 
as a function of energy. This approach has several advantages; in particular,
in this approach the only background of interest is that able to mimic the
signature, i.e. able to account for the whole observed modulation amplitude and to 
simultaneously satisfy all
the numerous specific peculiarities of the signature. No background of this sort has been 
found or suggested by anyone.

Thus, the DM annual modulation model-independent 
approach does not require any identification of S$_0$ from the total {\it single-hit} counting rate, 
in order to establish the presence of DM particles in the galactic halo. 

S$_0$ can be worked out only in a model dependent way, once S$_m$ has been determined. This 
procedure is performed by our collaboration by employing, within each specific framework, a 
maximum likelihood analysis which also takes into account the energy behaviour of each detector
(see literature).

In conclusion, the DM annual modulation signature allows
one to overcome the large uncertainties associated to the exploitation of
many data selections/subractions/statistical-discrimination 
procedures, to the modeling of surviving
background in keV region and to the {\it a priori} assumption on the nature and interaction type of 
the DM particle(s). In particular, as already mentioned e.g. in \cite{perflibra}, a precise 
modeling of background
in the keV region counting rate is always unlikely because e.g. of (i) the limitation of Monte-Carlo
simulations at very low energies; (ii) the fact that often just upper limits for residual
contaminants are available (and thus the real amount is unknown); (iii) the
unknown location of each residual contaminant in each component of the set-up;
(iv) the possible presence of non-standard contaminants, generally unaccounted;
(v) etc..

A visual indication has, however, been given in ref. \cite{taupnoz}, where 
the cumulative energy spectrum over the 25 detectors
has been reported up to 10 keV just as an example, showing that there was room
for a sizeable constant part of the signal: namely $S_0 < 0.25$ cpd/kg/keV in the [2,4] keV energy interval;
this has been discussed by our collaboration many times in presentations 
at conferences and workshops. 

As a matter of fact, any attempt to extrapolate a modeling of the 
background in the keV {\it single-hit} 
counting rate can only demonstrate that room for $S_0$ can exist, while it is not able to safely exclude 
a Dark Matter contribution. 

In conclusion, this explains why the conclusions derived from some fitting procedures used in ref. 
\cite{scem2} are untenable (see also later).

\section{Other additional comments}

Other comments deserve to be reported in the following.

\vspace{0.3cm}
{\bf \noindent Item regarding the so-called critique 1.}
We stressed in \cite{replay1} and references therein that the electron capture of $^{40}$K to the ground state of $^{40}$Ar
-- although its branching ratio is not well known from the theoretical and experimental points
of view -- provides a small (10\%) contribution to the total $^{40}$K contribution
at low energy in the {\it single-hit} counting rate. Moreover, the calculation of $^{40}$K
is not fully correctly performed in \cite{scem,scem2}, since e.g. the percentage of
K shell electron capture is not taken into account. In particular, the probabilities of
K shell electron capture to the first excited level (at 1461 keV) of $^{40}$Ar
(76.3\% \cite{tab_ref}) and to the ground state of $^{40}$Ar (87.9\% \cite{tab_ref}) 
must be included in the calculation.
Furthermore that small (10\%) contribution to the total $^{40}$K contribution
at low energy in the {\it single-hit} counting rate has always been included by our collaboration 
in all the evaluations \cite{damaweb}.

\vspace{0.3cm}
{\bf \noindent Item regarding the so-called critique 2.}
The authors in \cite{scem2} seem to forget that the DAMA/LIBRA set-up is made of 25 detectors,
each one with its own characteristics and energy spectrum. In the DAMA literature it was stated that
``The analysis has given for the $^{nat}$K content in the crystals values not exceeding 
about 20 ppb'' \cite{perflibra}; as is evident, 
this was not an upper limit on $^{nat}$K (with a relevant confidence level), but it is the 
maximum value among 
those measured in the 25 detectors. The average value (13 ppb)
has been published in ref. \cite{taupnoz} and discussed in many conferences.
Therefore, there are in \cite{scem,scem2} many statements (``no data is presented to support this number,
nor does the collaboration provide the uncertainty associated with it''; ``the lack of details so far provided
by the collaboration is unsettling.'', ...) not justified and unfounded.

\vspace{0.3cm}
{\bf \noindent Item regarding the so-called critique 3.}
A part from the arguments reported in Sect. 1 about the role of the modeling of the 
background components in the {\it single-hit} counting rate, we add here other specific 
comments.

The authors of ref. \cite{scem2} use different fitting functions (two segments 
$+$ a gaussian) and different parameters with respect to our fit and their choice appears
arbitrary.
In particular, all the considerations in ref. \cite{scem2} are also based on two questionable points:
\begin{enumerate}
\item the assumption that the background in the low energy region is flat;
\item the {\it a priori} belief that the level of this flat background in the [2,7] keV region can be arbitrarily 
fixed without taking into account any $S_0$ contribution from the Dark Matter signal\footnote{This 
assumption is always methodologically incorrect (and, in the 
particular case, also experimentally contradicted 
by the measured observable $S_m$ and by the absence of processes able to mimic it).}.
\end{enumerate}

As regards the first point, this assumption is justified in ref. \cite{scem2} with ``the 
universal feature of $\beta^-$ decays for small electron velocities'', but this motivation is not 
sufficient. In fact, other components can contribute to the counting rate in the 
low energy region, and
this is also evident in the given experimental spectrum, that is far from being flat in the [2,10] keV
region, as well as in the energy spectra published by other activities as shown e.g. in \cite{taupnoz}. 
Moreover, it is odd that the authors (of ref. \cite{scem2}) themselves violate their assumption by using a rising segment
to describe the behaviour in the [7,10] keV low energy region.
Different reasonable fits can provide slightly different upper limits for $S_0$, but the claim of a 0.85 
cpd/kg/keV
flat background in the [2,7] keV energy region is completely arbitrary because it is not based 
on the knowledge of the background contributions but it is the result of a fitting procedure
-- among others -- based on incorrect hypotheses.

The authors of \cite{scem2} conclude that a large modulation fraction ($S_m/S_0$) is required, but 
this is just the obvious consequence of their incorrect approach. 

\vspace{0.3cm}

Finally, just for completeness, let us note that in any case scenarios and Dark Matter 
candidates exist which can provide relatively large modulation fraction (see f.i. 
\cite{Freese12}).

\end{document}